\documentclass[a4paper]{article}   
\usepackage{epsfig}

\begin{document}                                                                \begin{center} 
{\LARGE Coherent Response in a Chaotic Neural Network}
\vspace{6mm} \\
{\large Haruhiko Nishimura${}^{1}$, Naofumi Katada${}^{1}$ and 
Kazuyuki Aihara${}^{2}$}  
\end{center} 
${}^{1}$Studies of Information Science, Hyogo University of Education, 942-1 Yashiro-cho, Hyogo 673-1494, Japan, E-mail:\{ haru, katada \}@life.hyogo-u.ac.jp\\
${}^{2}$The University of Tokyo, 7-3-1 Hongo, Bunkyo-ku, Tokyo 113, Japan, and\\
CREST, Japan Science and Technology Corporation(JST), 4-1-8 Hon-cho, \\
Kawaguchi, Saitama 332, Japan, E-mail:aihara@sat.t.u-tokyo.ac.jp

\begin{abstract}
We set up a signal-driven scheme of the chaotic neural network with the coupling constants corresponding to certain information, and investigate the stochastic resonance-like effects under its deterministic dynamics, comparing with the conventional case of Hopfield network with stochastic noise. It is shown that the chaotic neural network can enhance weak subthreshold signals and have higher coherence abilities between  stimulus and response than those attained by the conventional stochastic model.
\end{abstract}
\vspace{2mm} 

\section{Introduction}

Stochastic resonance(SR) is known as a phenomenon in which the presence of noise helps a nonlinear system in amplifying a weak (under barrier) signal~\cite{foo1}. The features of systems exhibiting SR seem to be applicable to some natural systems such as sensory neurons, which are noisy and operate as threshold systems. Since SR produces an information-transmitting phenomenon, its positive role in the neuronal processes is to be revealed. In fact, a single neuron model described by the FitzHugh-Nagumo (FHN) equations exhibits SR behavior~\cite{foo2} and this SR effect is found in the real sensory neurons located in the tail fan of crayfish~\cite{foo3}. The FHN equations driven by white noise and an arbitrary aperiodic signal are also examined in the context of excitable systems with threshold dynamics~\cite{foo4,foo5}. One can expect that the SR effect will be more pronounced in an ensemble of systems than in a single system. In view of a collective response of globally coupled bistable systems to periodic forcing, a neural network with dynamics of the Hopfield type~\cite{foo6} is studied,  under the assumption that white noise and the periodic signal are identical for all neurons~\cite{foo7}.

Given the basic three ingredients, that is, a form of threshold, a source of noise and a weak input signal, SR can generally be observed in a large variety of systems. Considering the fact that deterministic chaos resembles the feature of noise and provides a source of fluctuation, we have a natural question whether SR-like behavior can be observed in deterministic dynamical systems in the absence of noise. Two different approaches to this problem are known. One way is to substitute the stochastic noise by a chaotic source. This situation in which the chaos is supplied as an additive noise, resembles the conventional setup for SR and yields SR-like enhancement as expected~ \cite{foo8,foo9}. The other approach is to use the intrinsic chaotic dynamics of a nonlinear map. No external source is necessary to provide the randomness. This method generates a sort of activated hopping process which is then synchronized by a weak periodic signal~\cite{foo10,foo11}.

Until now, diverse types of chaos have been confirmed at several hierarchical levels in the real neural systems from single cells to cortical networks (e.g. ionic channels, spike trains from cells, EEG)~\cite{foo12}. By producing chaos as effective noise spontaneously, biological systems may enhance their functions through signal amplification. This scenario is likely to have occurred even in the associative memory dynamics. The chaotic neural network model~\cite{foo13} is known as a framework beyond the Hopfield neural network~\cite{foo6} with only equilibrium point attractors. Its dynamic retrieving and learning features have been studied~\cite{foo14,foo15,foo16}. In this paper, we set up a signal-driven scheme of the chaotic neural network with the coupling constants corresponding to certain information, and investigate the SR-like effects under its deterministic dynamics, comparing with the conventional case of the Hopfield network with stochastic noise.

\section{Models}

For N neurons connected by synaptic couplings $w_{ij}$ with $w_{ii}=0$, the system (signal-driven scheme) is described by
\begin{eqnarray}
X_i(t+1) = f(h_i(t)+S_i(t))~~,
\end{eqnarray}
where $X_{i}$: output of neuron $i(-1 \leq X_i \leq 1)$, $w_{ij}$: synaptic weight from neuron $j$ to neuron $i$, $f$: output function defined by $f(y)=tanh(y/2\varepsilon)$ with the steepness parameter $\varepsilon$, $h_i$: internal potential, $S_i$: contribution of external input signal. This form is a simple and possible incorporation of stimuli as the changes of neuronal activity.
In the case of chaotic neural network (CNN)~\cite{foo13}, the internal potential is given by
\begin{eqnarray}
h_i(t) &=& \eta_i(t)+\zeta_i(t)~~,\\
\eta_i(t) &=& \sum_{j=1}^N w_{ij} \sum_{d=0}^{t} k_{f}^{d} X_{j}(t-d)~~,\\ 
\zeta_i(t) &=& - \alpha \sum_{d=0}^{t}k_{r}^{d} X_{i}(t-d) - \theta_{i}~~,
\end{eqnarray}
where $\theta_{i}$: threshold of neuron $i, k_{f}(k_{r}$): decay factor for the feedback (refractoriness) $(0 \leq k_{f},~k_{r} <1),~\alpha$: refractory scaling parameter. Owing to the exponentially decaying form of the past influence, the dynamics of $\{ \eta_{i} \}$ and $\{ \zeta_{i} \}$ can be described as follows :
\begin{eqnarray}
\eta_{i}(t) &=& k_{f} \eta_{i}(t-1) + \sum_{j=1}^{N} w_{ij} X_{j}(t) ~~,\\
\zeta_{i}(t) &=& k_{r} \zeta_{i}(t-1) - \alpha X_{i}(t) - \theta_{i}(1-k_{r})~~. 
\end{eqnarray}
When $\alpha=k_{f}=k_{r}=0$, the network corresponds to the conventional discrete-time Hopfield network (we call the Hopfield network point (HNP)):
\begin{eqnarray}
X_i(t+1) = f\Bigl(\sum_{j=1}^N w_{ij} X_j(t) - \theta_i\Bigl)~~.
\end{eqnarray}
We also look into the case that stochastic fluctuations are attached to HNP in Eq.(7):
\begin{eqnarray}
h_i(t) = \sum_{j=1}^{N} w_{ij} X_{j}(t) - \theta_{i} + F_{i}(t)~~,
\end{eqnarray}
where $F_{i}(t)$ is a neuron-independent Gaussian white noise defined by $<F_{i}(t)>=0$ and $<F_{i}(t) F_{j}(t')> = D^{2} \delta_{t, t'} \delta_{i, j}$, and $D$ is the noise intensity parameter.We call Eq.(8) a stochastic neural network(SNN).

The synaptic configuration $\{ w_{ij} \}$ is determined by storing pattern information in the network as minima of the computational energy :
\begin{eqnarray}
E=-\frac{1}{2}\sum_{ij}w_{ij}X_iX_j
\end{eqnarray}
at HNP. This is done by using a local iterative learning rule~\cite{foo17} for $p$ patterns $\{ \xi_i^\mu\} \equiv (\xi_1^\mu, \cdots, \xi_N^\mu)$, $(\mu=1, \cdots, p$; $\xi_i^\mu= +1~~or~-1)$ in the following form :
\begin{eqnarray}
w_{ij}^{new} = w_{ij}^{old} + \sum_\mu\delta w_{ij}^\mu
\end{eqnarray} 
with
\begin{eqnarray}
\delta w_{ij}^\mu=\frac{1}{N}\theta (1-\gamma_i^\mu)\xi_i^\mu\xi_j^\mu ,
\end{eqnarray}
where $\gamma_i^\mu \equiv \xi_i^\mu \sum_{j=1}^Nw_{ij}\xi_j^\mu$ and $\theta(h)$ is the unit step function.

\section{Simulations and Results}

To carry out computational experiments, we consider a network with $N=156$, $\{ \theta_{i} \}=0$ and $\varepsilon=0.015$(unless otherwise stated), and use non-orthogonal 20 random  patterns $R^1 \sim R^{20}$ as a set of external signal: $ \{ S_i \} = s \{ \xi_{i}^{\mu} \} (\mu=1, \cdots, 20)$. $s$ is the strength factor of signal. 
$\{ \xi_{i}^{\mu} \}$ ($i=1, \cdots, N$) is represented by $12 \times 13$ binary data and a half of $\xi_i^\mu$'s has $+1$ and the other half has $-1$. $10$ patterns $R^{1} \sim R^{10}$ are stored with the above learning rule of Eq.(10) ($p=10$), and then the corresponding multi-stable landscape is made on the network dynamics.

\begin{figure}
\centerline{\epsfig{file=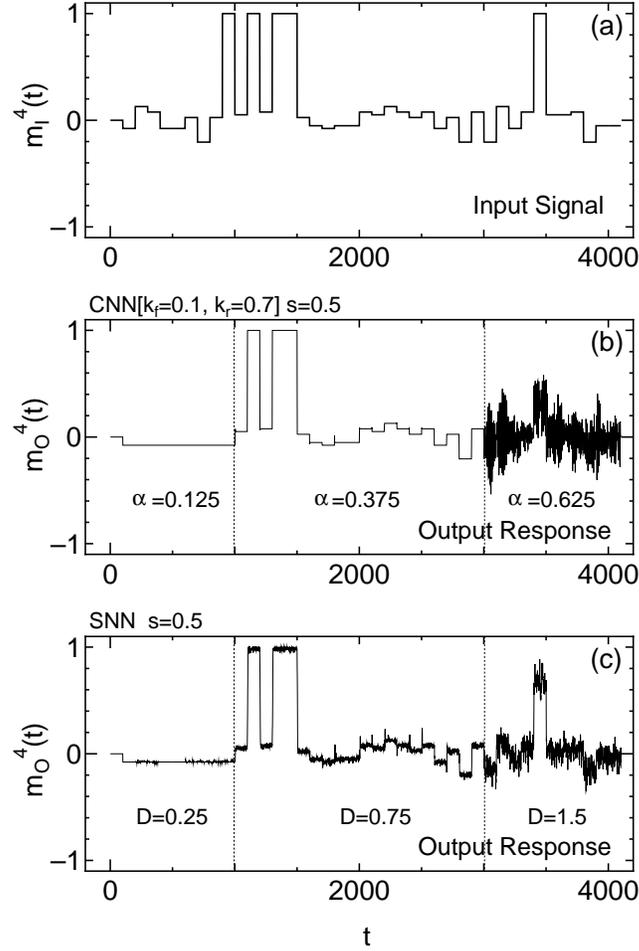,width=8.5cm}}
\caption{(a) Time series of the input signal $\{ S_{i}(t) \}$ with $T_{I}=100$ used in our simulations. (b) The behaviors of the output response $\{ X_{i}(t) \}$ in CNN with different values of the refractory parameter $\alpha$ for the signal ($s$=0.5). (c) Those in SNN with different values of the noise intensity parameter $D$. In (a) $\sim$ (c) the ordinate is the relative overlaps with the pattern $R^{4}$ [Eqs.(12),(13)].}
\end{figure}

As the temporal input signal $\{ S_{i}(t) \}$, we take two kinds of signals. One is a random train composed of stored patterns $R^{1} \sim R^{10}$ (e.g. $[R^{3} \rightarrow R^{7} \rightarrow R^{2} \rightarrow \cdots]$) with a duration $T_{I}$ for every pattern and the other is a similar sequence of non-stored patterns $R^{11} \sim R^{20}$. Figure $1$ shows typical examples of the results of temporal behaviors of the neural networks to input signal $\{ S_{i}(t) \}$. An input random train of stored patterns ($T_{I}=100$) is given in Fig.1(a) by the relative overlap with pattern $R^4$:
\begin{eqnarray}
m_I^4(t) = \frac{1}{N} \sum_{i=1}^N \tilde{S}_{i}(t) \xi_i^4~~,~~\tilde{S}_i=S_{i}/s~~. 
\end{eqnarray}
Responses of CNN ($k_f=0.1$, $k_r=0.7$) and SNN to this signal with strength $s=0.5$ are shown in Fig.1(b) and (c) by
\begin{eqnarray}
m_0^4(t) = \frac{1}{N} \sum_{i=1}^N X_i(t) \xi_i^4~~. 
\end{eqnarray}
In CNN [Fig.1(b)], as we can see from three cases where $\alpha$ is $0.125$, $0.375$ and $0.625$, the performance of output response is largely affected by its refractory scale parameter $\alpha$. The response fits well the input signal when $\alpha$ is $0.375$. From Fig.1(c), similar behaviors are observed for SNN against the noise intensity parameter $D(=0.25, 0.75$ and $1.5)$, except for the appearance of noise-driven fluctuations.

To evaluate the coherence between the signal \{$S_i(t)$\} and the response \{$X_i(t)$\}, we introduce the correlation coefficient $r$ and the discrimination efficiency $n$. These quantities are defined as
\begin{eqnarray}
r=\frac{\overline{Dev(m_I^4)Dev(m_0^4)}}{[\overline{Dev(m_I^4)^2}]^{1/2}[\overline{Dev(m_0^4)^2}]^{1/2}}
\end{eqnarray}
and
\begin{eqnarray}
n=\overline{\frac{1}{N} \mathop{\sum}_{i=1}^N \tilde{S}_{i}(t) X_i(t)}~~,~~\tilde{S}_i=S_{i}/s~~,
\end{eqnarray}
where $Dev(Y)$ is the deviation $Y-\overline{Y}$ and the overbar denotes an average over time. $n$ naively means the rate of information transfer from stimulus to response.

\begin{figure}
\centerline{\epsfig{file=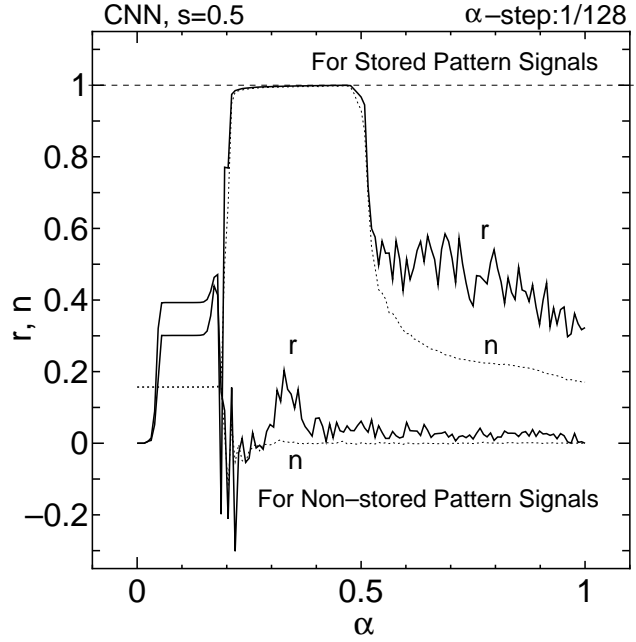,width=8.5cm}}
\caption{Correlation coefficient $r$ (solid line) and discrimination efficiency $n$ (dotted line) as a function of refractory parameter $\alpha$ in CNN simulations with Eq.(2). The results for the input signals of stored and non-stored pattern trains are  shown.}
\end{figure}

The numerical results of $r$ and $n$ plotted against refractory factor $\alpha$ are given in Fig.2, calculated using the temporal data at every different $\alpha$ values. The results for $\{ S_{i}(t) \}$ of stored patterns $R^{1} \sim R^{10}$ are in contrast with the results for $\{ S_{i}(t) \}$ of non-stored patterns $R^{11} \sim R^{20}$ in the range of $\alpha$ larger than about $0.2$ wherein the response is hopping out and into a well (minimum of the energy) corresponding to a stored pattern state. This movability causes an increased coherence with stored pattern stimuli. Under the same conditions of input signal, dependences of $r$ and $n$ on the noise intensity $D$ in SNN are examined as shown in Fig.3. Between stored and non-stored pattern signals, a quite difference like in CNN  is also confirmed. Comparing the results for stored pattern signal, we can see that in CNN there appears a projected plateau for both $r$ and $n$, and in this flat region both of their values are kept very close to 1. Contrary to this, in SNN $r$ and $n$ have gradual variations with $D$ and $n$ degrades faster than $r$, which seems to be consistent with conventional SR. This comparison indicates that CNN can cause SR-like phenomena with high performance which cannot be attained in SNN.

\begin{figure}
\centerline{\epsfig{file=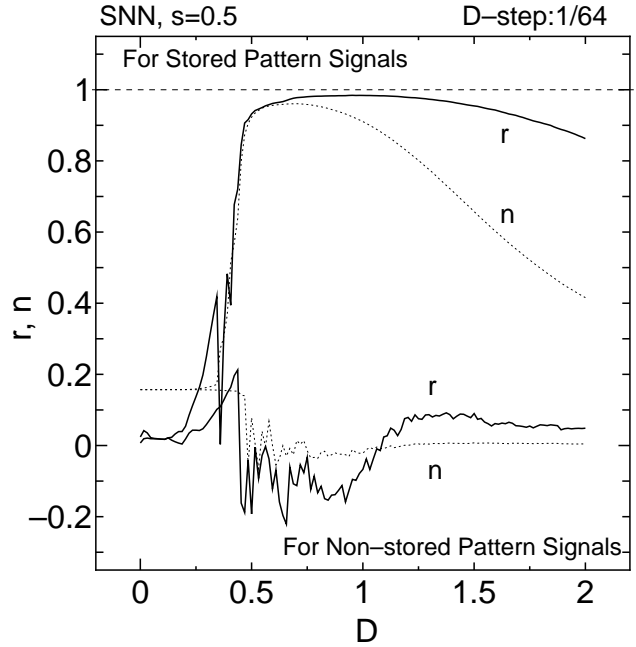,width=8.5cm}}
\caption{$r$ (solid line) and $n$ (dotted line) as a function of noise intensity parameter $D$ in SNN simulations with Eq.(8).}
\end{figure}

In CNN, the stronger the signal strength $s$ is, this $\alpha$-range of the flat plateau becomes wider. Conversely, this range becomes narrower for weaker $s$. We illustrate this effect as coherent $\alpha$-range versus signal strength $s$ in Fig.4, together with the relationship of maximum Lyapunov exponent $\lambda_1$~\cite{foo18} to $\alpha$ when there exists no signal forcing. As $s$ decreases, the $\alpha$-range ($r \geq 0.9$) becomes narrow and at last disappears. Then the $\alpha$ value coincides with  $\alpha_\ast$ at the sudden rise point of the Lyapunov exponent $\lambda_1$. This fact tells  that sensitive and flexible responses happen around the boundary between order and disorder, in other words, the edge of chaos~\cite{foo19}. At $\alpha < \alpha_\ast$, the refractory term ($-\alpha X_{i}$) in Eq.(6) makes a well (minimum of the energy) shallow in effect and helps the input signal drive the network state. On the other hand, at $\alpha > \alpha_\ast$ the chaotic attractor is driven by the input signal and stabilized to the corresponding network state.

Figure 5 shows return maps of the internal potential $h_{i}$ for a neuron ($i=12$) in CNN with no signal forcing ($s=0$) when $\alpha > \alpha_\ast$. In the case of $\alpha=58/128$, the trajectories of $h_{i}$ are attracted into the region of square ($-0.5 \sim 0.5$) such that controllable by the input signal $s=0.5$. When $\alpha=96/128$ (a deeper chaotic state), however, the trajectories are attracted outward so that chaos cannot be suppressed by the input signal $s=0.5$.

We have investigated all the above phenomena in other conditions of the parameters and the input signals, and have found similar results and overall tendency.

\begin{figure}
\centerline{\epsfig{file=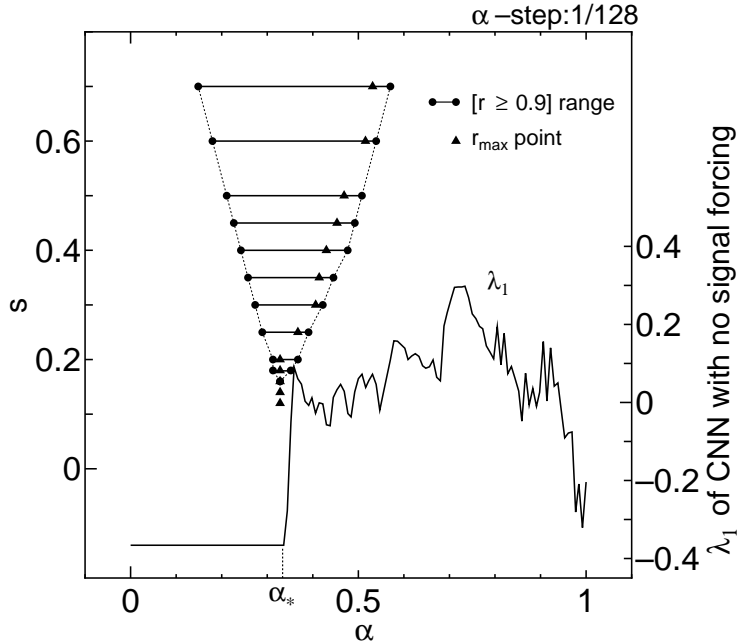,width=10.0cm}}
\caption{Dependence of the coherent $\alpha$-range ($r \geq 0.9$) and the $r_{max}$ point on the input signal strength $s$ (the left-axis) in CNN, shown together with the maximum Lyapunov exponent $\lambda_1$ (the right-axis) as a function of parameter $\alpha$ in CNN with no signal forcing($s=0$).}
\end{figure}

\begin{figure}
\centerline{\epsfig{file=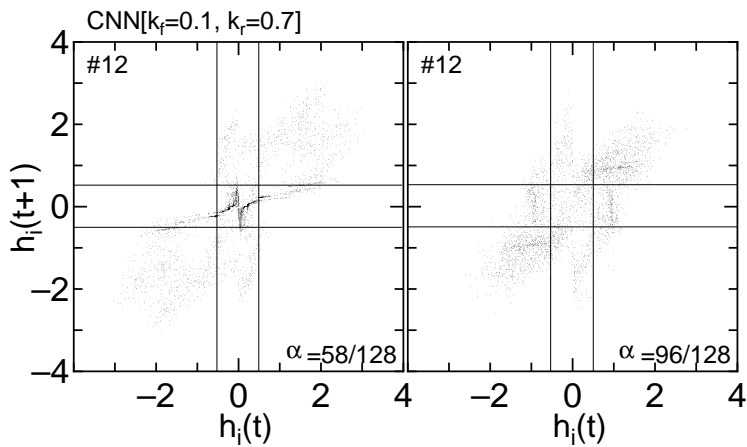,width=10.0cm}}
\caption{Return maps of the internal potential for the 12th neuron ($i=12$) in CNN with no signal forcing in the cases of $\alpha=58/128$ and $96/128$ ($>\alpha_\ast$).}
\end{figure}

\section{Conclusion}

We have shown that the chaotic neural network can enhance weak subthreshold signals and have higher coherence abilities between  stimulus and response than those attained by the conventional stochastic neural network model. The high coherent response is found to arise around the edge of chaos. This implies that some of SR phenomena may be realized by the inherent properties of deterministic nonlinear systems without any external noise. Analytical study to explain these results will be important in our future work. The coherent response concept is expected to be related to researches in cognitive neuroscience from dynamical system viewpoints~\cite{foo20,foo21}.

\end{document}